\documentclass[11pt,a4paper]{article}
\usepackage{jheppub}
\usepackage{amsmath}
\usepackage[most]{tcolorbox}
\usepackage{dsfont}
\usepackage{ulem}
\usepackage{natbib}
\usepackage{xcolor}
\usepackage[hang,flushmargin]{footmisc}
\usepackage{tikz-cd}
\usepackage{enumitem}
\usepackage{amsfonts,amssymb, amscd,amsmath,latexsym,amsbsy,bm,amsthm}
\usepackage{stmaryrd}
\usepackage{todonotes}
\usepackage{float}
\usepackage{tikz}
\usetikzlibrary{positioning}

\usepackage{nameref}

\usepackage{romannum}

\makeatletter\renewcommand{\@biblabel}[1]{#1.}\makeatother

\newtcolorbox{empheqboxed}{colback=gray!20, 
 colframe=white,
 width=\textwidth,
 sharpish corners,
 top=0mm, 
 bottom=0pt
}

\title{Bailey chain approach to 2d 
\texorpdfstring{$\mathcal{N}=(0,2)$}{N=(0,2)}
 dualities
}
\author{Zehra Akbulut$^{a}$, Ilmar Gahramanov$^{a,b}$, Anıl Kahraman$^{a,c}$, Mustafa Mullahasanoglu$^{a,d}$ and Yaren Yıldırım$^{a,c}$}
\affiliation{
$^a$Department of Physics, Bogazici University, 34342 Bebek, Istanbul, Türkiye\\[-0.4cm]

$^b$Center for Mathematics and its Applications, Khazar University, \\ Mehseti St. 41, AZ1096, Baku, Azerbaijan\\[-0.4cm]

$^c$Department of Mathematics, Bogazici University, 34342 Bebek, Istanbul, Türkiye\\[-0.4cm]

$^d$Feza Gursey Center for Physics and Mathematics, Bogazici University,\\ 34684, Kandilli,
Istanbul, Türkiye
}

\emailAdd{zehra.akbulut@std.bogazici.edu.tr}
\emailAdd{ilmar.gahramanov@bogazici.edu.tr}
\emailAdd{anil.kahraman@std.bogazici.edu.tr}
\emailAdd{mustafa.mullahasanoglu@std.bogazici.edu.tr}
\emailAdd{yaren.yildirim@std.bogazici.edu.tr}

\abstract{ We study a two-dimensional $\mathcal{N}=(0,2)$ supersymmetric duality and construct novel Bailey pairs for the associated elliptic genera. 
This framework provides a systematic method to establish the equivalence of the elliptic genera of quiver gauge theories generated via iterative applications of the seed duality. 
}

\begin{document}

\maketitle

\section{Introduction}

In recent years, numerous supersymmetric dualities \cite{Gadde:2013lxa, Dedushenko:2017osi,Sacchi:2020pet,Jiang:2024ifv,Amariti:2024usp,Amariti:2025jvi} for two-dimensional $\mathcal{N} = (0,2)$ theories have been developed, beginning with the work \cite{Gadde:2013lxa} that identified a new IR equivalence. 

Bailey's lemma is one of the most powerful tools for systematically generating hypergeometric identities, and its framework has also found applications in computations of supersymmetric gauge theories \cite{Kashaev:2012cz,Gahramanov:2015cva,Brunner:2017lhb,Gahramanov:2021pgu,Gahramanov:2022jxz,Catak:2022glx, Catak:2024ygo}. In this work, we focus on a specific IR duality and construct a new Bailey tree for the associated elliptic genera. This construction provides a systematic method for establishing the equivalence of the elliptic genera \cite{Benini:2013nda,Benini:2013xpa} of quiver gauge theories generated via iterative applications of the seed duality.

We construct Bailey pairs for the following two-dimensional $\mathcal{N} = (0,2)$ supersymmetric dual theories \cite{Dedushenko:2017osi,Sacchi:2020pet}:

\paragraph{Theory A:}
An $SU(2)$ gauge theory with four chiral multiplets $Q_a$ ($a = 1, \dots, 4$)  in the fundamental representation of the
gauge group and without superpotential.

\paragraph{Theory B:}
Landau--Ginzburg model with one Fermi multiplet $\Psi$ and six chiral multiplets $\Phi_{ab} = -\Phi_{ba}$ for $a,b = 1, \dots, 4$, interacting via the cubic superpotential\footnote{Here $\mathrm{Pf}(\Phi)$ stands for the Pfaffian of the antisymmetric $4\times4$ matrix $\Phi_{ab}$, namely, $\mathrm{Pf}(\Phi) = \Phi_{12}\Phi_{34} + \Phi_{13}\Phi_{42} + \Phi_{14}\Phi_{23}.$} $ W = \Psi\, \mathrm{Pf}(\Phi)$. 

This provides a two-dimensional $\mathcal{N} = (0,2)$ analogue of the $\mathcal N=1$ four-dimensional Seiberg duality between $SU(2)$ gauge theory with four flavors and the Pfaffian superpotential theory. The equality of elliptic genera for the two dual theories implies the following
$q$-hypergeometric integral identity
\begin{align} 
\frac{(q;q)_\infty^2}{2}\oint \frac{\theta(z^{\pm2})} {\prod_{i=1}^4\theta(a_i z^{\pm1})} \frac{dz}{2\pi i z}
=\frac{\theta\left(q/\prod_{i=1}^4a_i\right)
          }{
          \prod_{1\leq i<j\leq 4}\theta(a_i  a_j)}  \label{decorationequation}\; ,
\end{align}
where there is a constraint on fugacities $\prod_{i=1}^4a_i=k^4$ and $\theta(z)\equiv \theta(z;q)$ is the theta function defined as
\begin{align}
    \theta(z;q)=(z;q)_\infty (qz^{-1};q)_\infty\:,\quad \text{with}\quad (z;q)_\infty =\prod_{i=0}^\infty(1-zq^i)\:.
\end{align}
We also use the shorthand notation $\theta(z^{\pm1})= \theta(z)\theta(z^{-1})$ throughout the paper.

In \cite{Sacchi:2020pet}, the following two-dimensional $\mathcal{N}=(0,2)$ duality for quiver gauge theories  was derived by dimensional reduction from the four-dimensional $\mathcal{N}=1$ duality:

\paragraph{Theory A:} 
An $SU(2)$ gauge theory containing four fundamental chiral multiplets 
$L_a, R_a$ ($a = 1, 2$) and two gauge-singlet Fermi multiplets 
$\Psi_L, \Psi_R$ with the superpotential $W = \Psi_L\, L \cdot L + \Psi_R\, R \cdot R.$

\paragraph{Theory B:}
A linear quiver with $N - 1$ $SU(2)$ gauge nodes connected by 
bifundamental chiral multiplets $Q_{(i,i+1)}$ for 
$i = 1, \ldots, N - 2$. 
There is one additional fundamental chiral multiplet at each end of the chain, 
$Q_{(0,1)}$ and $Q_{(N-1,N)}$, as well as $N$ gauge-singlet 
Fermi multiplets $\Psi^{(i)}$ ($i = 1, \ldots, N$). The superpotential in this theory  takes the form $W =    \sum_{i=1}^{N-1} 
    \Psi^{(i)}\, Q_{(i-1,i)}\, Q_{(i,i+1)}$.

In \cite{Sacchi:2020pet}, by repeatedly applying the integral identity (\ref{decorationequation}), the author obtained the following integral identity corresponding 
to the duality for an arbitrary value of $N$
\begin{equation}
\begin{aligned}
\frac{(q;q)_\infty^{2N}}{2^N}
\oint \frac{\prod_{i=1}^{N}\theta(s_i^2) }{     \prod_{i=1}^{N} \theta(s_iz_{i-1}^{\pm1} z_i^{\pm1})}    \prod_{i=1}^{N-1}\frac{\theta(z_i^{\pm2}) \mathrm{d}z_i}{2\pi i z_i}
= 
\frac{\theta\left(q/\prod_{i=1}^{N}s_N^{2}\right)
          }{
 \theta( \prod_{i=1}^{N}s_iz_0^{\pm1}z_N^{\pm1})}\:.
 \label{N-integral}
\end{aligned}
\end{equation}

The main point of this paper is that the above integral identity 
can be obtained through the Bailey tree construction starting from 
the integral identity~(\ref{decorationequation}).  In other words, the repeated application of the identity can be  interpreted as successive applications of the Bailey transformation.

\section{Bailey pairs}
In this section, we closely follow the Bailey construction presented in~\cite{Catak:2024ygo}.

We define an integral operator $M(t)_{z; x}$, consisting of an integration over $x \in \mathbb{C}$. 
The operator acts as a transformation between Bailey pairs, mapping 
$\alpha(x; t)$ to a new function $\beta(z; t)$ with respect to the parameter $t\in \mathbb{C}$
\begin{equation}\label{firstBaileypair}
    \beta(z;t) = M(t)_{z;x}\alpha(x;t) \;.
\end{equation}

We introduce an operator $I(s)$, where $s \in \mathbb{C}$, 
which acts by redefining the variables in a given function. 
This operator satisfies the following properties
\begin{equation}
    I(s)\, I(s^{-1}) = 1, 
    \qquad 
    I(1) = 1, 
    \qquad 
    I(t)\, I(s) = I(st).
\end{equation}
We also define new functions generated from the Bailey pairs
\begin{align}\label{defining}
    \alpha'(x;ts) &=I(s) \alpha(x; t) \:,\\
    \beta'(x; ts)&=I(s^{-1}) M(s)_{x;z} I(ts)\beta(z; t)\:.
\end{align}
We require these new functions to form a Bailey pair~(\ref{firstBaileypair}) with respect to the parameter $ts$
    \begin{equation}
    \beta'(w;ts) = M(ts)_{w;x}\alpha'(x;ts)\:. 
\end{equation}
To ensure this property, the operator $M$ must satisfy the following relation\footnote{This relation is also referred to as the decoration transformation in the context of the gauge/Yang–Baxter Equation correspondence, see \cite{Catak:2024ygo, Catak:2025fye, Mullahasanoglu:2023nes}.}
\begin{equation}
\label{operatordecoration}
\begin{gathered}
       M(s)_{w; z}  M(t)_{z; x} =  M(st)_{w; x}  \:.
\end{gathered}
\end{equation}
When this relation holds, it follows that a sequence of functions can be constructed via the integral operator, a construction referred to as the Bailey lemma.  The $M$ operator with non-trivial $D$ operator constructed in \cite{Gahramanov:2022jxz} satisfies the star-triangle relation, which is a version of the Yang-Baxter equation.

Here, we introduce the integral operator $M$ to construct Bailey pairs for the equality of elliptic genera for the dual theories $(\ref{decorationequation})$, all necessary ingredients for this duality have been mentioned above,
\begin{equation}
\begin{aligned}
M(t)_{z;x}f(x)=
\theta(t^{-2})\frac{(q;q)_\infty^{2}}{2}\oint
\frac{ \mathrm{d}x}{2\pi i x}
\frac{\theta(x^{\pm2})}{        \theta(t^{-1}z^{\pm1} x^{\pm1})} f(x)\:.
\end{aligned}\label{SU2Moperator}
\end{equation}
One can show that the operator~\eqref{SU2Moperator} satisfies the relation~\eqref{operatordecoration} 
by making use of the integral identity~\eqref{decorationequation} together with the following change of variables
\begin{equation}
\begin{aligned}
    a_{1,2} &= s^{-1} w^{\pm1}, 
    \qquad 
    a_{3,4} = t^{-1} x^{\pm1}.
    \label{changeofvariables1}
\end{aligned}
\end{equation}
The key result of this work is that the integral identity (\ref{N-integral}) is a natural consequence of the Bailey tree. When one generates $N^{th}$ function in the sequence constructed by the Bailey Lemma, the iterative application of the integral operator should satisfy the following relation 
\begin{equation}
\begin{gathered}
      \prod_{i=1}^{N} M(s_i)_{z_{i-1}; z_{i}}     =
       M\left(\prod_{i=1}^{N}s_i\right)_{z_{0}; z_{N}}  \:.
\end{gathered}
\end{equation}
We substitute the explicit expression for the integral operator (\ref{SU2Moperator}), which satisfies relation (\ref{decorationequation}), and thereby directly obtain 
\begin{equation}
\begin{aligned}
\frac{(q;q)_\infty^{2N}}{2^N}\oint \frac{\prod_{i=1}^{N}\theta(s_i^2) }{     \prod_{i=1}^{N} \theta(s_iz_{i-1}^{\pm1} z_i^{\pm1})}    \prod_{i=1}^{N}\frac{\theta(z_i^{\pm2}) \mathrm{d}z_i}{2\pi i z_i}
= 
\oint \frac{\theta\left(q/\prod_{i=1}^{N}s_i^{2}\right)
          }{
 \theta( \prod_{i=1}^{N}s_iz_0^{\pm1}z_N^{\pm1})}\frac{\theta(z_N^{\pm2}) \mathrm{d}z_N}{2\pi i z_N}\:.
\end{aligned}
\end{equation}
Thus, we establish that iterating the integral identity is equivalent to the sequential application of the defining relation for the Bailey tree.

We now construct new Bailey pairs generated from an initial pair~\eqref{firstBaileypair}. 
Recalling that $M(t)_{z;x}$ is an integral operator acting on a sequence of functions $f(x)$, 
the relation~\eqref{firstBaileypair} suggests starting with the seed function
\begin{equation}
    \alpha(x;t) = \delta(x - u),
\end{equation}
where $u \in \mathbb{C}$ denotes a new complex parameter.

Then the function $\beta(z;t)$ of the following form
\begin{equation}
\begin{aligned}
    \beta(z;t) &= M(t)_{z;x}\, \delta(x - u) \\
               &:= M(t; z; u),
\end{aligned}
\end{equation}
forms a Bailey pair together with $\alpha(x;t)$. From this seed, we can generate new Bailey pairs using the Bailey lemma
\begin{align}
    \alpha(x; ts) &= I(s)\, \alpha(x;t), \\
    \beta(x; ts)  &= I(t^{-1})\, M(s)_{x;z}\, I(st)\, \beta(z;t).
\end{align}

The relation $\eqref{firstBaileypair}$ does not give us a particularly interesting result, as it yields the integral identity (\ref{decorationequation}), which we have used to prove the Bailey lemma
\begin{equation}
\begin{aligned}\
M(s)_{w;z}I(st)M(t;z;u) = I(t)M(st ; w; u)I(s)\label{secondbaileydef}  \;.
\end{aligned}
\end{equation}
An immediate consequence of $\eqref{secondbaileydef}$ is the functions $\Tilde{\alpha}(z;s)$ and $\Tilde{\beta}(w;s)$ defined by
\begin{align}
\Tilde{\alpha}(z;s) &= I(st)M(t;z;u)\:, \\
\Tilde{\beta}(w;s) &=I(t)M(st ; w; u)I(s) \;,
\end{align}
form a Bailey pair with respect to parameters $s \in \mathbb{C}$. Applying the lemma once again, we find
\begin{align}
\Tilde{\alpha}'(z;sp) =& I(p)I(st)M(t;z;u)\:, \\
    \begin{split}
    \Tilde{\beta}'(x;sp) =& I(s^{-1})M(p)_{x;w}I(sp) I(t)M(st ; w; u)I(s)  \;, 
    \end{split}
\end{align}
where $u, p \in \mathbb{C}$ are arbitrary. The relation
\begin{equation}
\Tilde{\beta}'(x;sp) = M(sp)_{x;z} \Tilde{\alpha}'(z;sp)   \;,
\end{equation}
yields a non-trivial integral identity
\begin{equation}
\begin{aligned}
M(p)_{x;w}I(sp)I(t)M(st ; w; u) = I(s^{-1})I(s) M(sp)_{x;z} I(p)I(st)M(t;z;u) \;.
\end{aligned}
\end{equation}
One can see the reduction 
\begin{equation}
\begin{aligned}
M(p)_{x;w}M(st ; w; u) = M(sp)_{x;z} M(t;z;u) \;,
\label{flippingrelation}
\end{aligned}
\end{equation}
where the equality is used as the flipping relation \cite{ Catak:2025hoz} in the context of the gauge/Yang–Baxter equation correspondence.

We need the function $M(t;z;u)$ due to the new definition of $\beta(z;t)$ after modifying  $\alpha(x;t) = \delta(x-u)$
\begin{equation}
\begin{aligned}
M(t;z;u)=\theta(t^{-2})
\frac{(q;q)_\infty^{2}}{2}
\frac{ \theta(u^{\pm2})}{        \theta(t^{-1}z^{\pm1} u^{\pm1})} \:.
\end{aligned}
\end{equation}
If one use the change of variables
\begin{equation}
\begin{aligned}
     a_{1,2}=p^{-1} x^{\pm1}, \quad  
a_{3,4}=t^{-1}s^{-1} u^{\pm1}, 
     \\
\Tilde{a}_{1,2}=p^{-1} s^{-1} x^{\pm1}, 
      \quad    \Tilde{a}_{3,4}=t^{-1} u^{\pm1} \:,\label{changeofvariables3}
\end{aligned}
\end{equation}
can quickly reach that the relation (\ref{flippingrelation}) holds by the integral transformation formula in terms of theta functions
\begin{align}
\frac{(q;q)_\infty^2}{2}\oint \frac{\theta(z^{\pm2})} {\prod_{i=1}^4\theta(a_i z^{\pm1})} \frac{dz}{2\pi i z}
=\frac{\theta(q/\Tilde{a}_1\Tilde{a}_2)\theta(\Tilde{a}_3\Tilde{a}_4)}{\theta(a_1a_2)\theta(q/a_3a_4)}
\frac{(q;q)_\infty^2}{2}\oint \frac{\theta(z^{\pm2})} {\prod_{i=1}^4\theta(\Tilde{a}_i z^{\pm1})}\frac{dx}{2\pi i x}
\label{flippingequation}\:,
\end{align}
where the constraint is $\prod_{i=1}^4a_i=k^8$ and we also redefine some parameters together with the modification with a chemical potential $s$
\begin{align}
    \Tilde{a}_{1,2}=a_{1,2}s \:,\quad
\Tilde{a}_{3,4}=a_{3,4}s^{-1} \:.
\label{tilde1}
\end{align}

The integral identity (\ref{flippingequation}) is can be obtained using the double integral method \cite{Catak:2021coz} by applying the integral identity (\ref{decorationequation}), so it is expected to have the same integral operator $M$.

\section{Conclusions}\label{conclusion}

In conclusion, we have studied a particular two-dimensional $\mathcal{N}=(0,2)$ supersymmetric duality and developed new Bailey pairs for the corresponding elliptic genera. This construction provides a systematic method to demonstrate the equivalence of the elliptic genera of quiver gauge theories obtained through iterative application of the seed duality.

Analogues of the Bailey pairs constructed here have found applications in knot invariants \cite{Kashaev:2012cz}, Yang–Baxter relations \cite{Gahramanov:2015cva, Gahramanov:2022jxz}, and other mathematical structures. Exploring similar constructions in these areas would be an interesting direction for future work.




\section*{Acknowledgements}

All authors are supported by Istanbul Integrability and Stringy Topics Initiative (\href{https://istringy.org/}{istringy.org}). Ilmar Gahramanov, Mustafa Mullahasanoglu, and Yaren Yıldırım are partially supported by the Scientific and Technological Research Council of Turkey (TÜBİTAK) under the grant number 122F451. Ilmar Gahramanov would like to thank the Nesin Mathematics Village (Şirince, Türkiye) for its hospitality, where part of this work was carried out. 
Mustafa Mullahasanoglu is also grateful to Masahito Yamazaki for his valuable discussions on this and related subjects and would like to thank him for the warm hospitality at the Kavli Institute for the Physics and Mathematics of the Universe and the University of Tokyo, where part of this work was carried out.



\bibliographystyle{utphys}
\bibliography{refYBE}

\end{document}